# VTracker: Impact of User Factors On Users' Intention to Adopt Dietary Intake Monitoring System with Auto Workout Tracker

[1]Oh Tai Lan, [2]Ahmad Suhaimi Baharudin, [3*]Kamal Karkonasasi

*School of Computer Science, Universiti Sains Malaysia (USM), Malaysia*

*asasi.kamal@gmail.com

**Abstract**

Nowadays, Malaysian are so concerned about their body health. In respond to this, this study proposed a conceptual prototype called vTracker to assist its users to have a healthier body. vTracker is a web-based mobile application which helps users to self-monitor their dietary intakes and workout activities in a few simple steps. A research framework has been proposed using the Unified Theory of Acceptance and Use of Technology (UTAUT) model to understand the level of users' intention to adopt vTracker in Malaysia. Data was collected from 206 respondents in Malaysia using survey method. Based on the result of the analysis, it was found that respondents agree to subscribe to the system. It addition to that, it was found that the three factors have a positive impact on users'intention to adopt vTracker. These mentioned factors are performance effort expectancy (PEE), social influence (SI) and facilitating condition (FC). These significant factors are used for designing vTracker portal.

## 1. Introduction

Although Malaysia is moving towards achieving vision 2020 as a developed nation, the percentage of overweight and obese population are still on the rise. Users no longer need to go through the inconveniences process to complete their routines. A conducted survey by The Lancet on year 2014 in Malaysia shows that 49% of women and 44% of men are obese. It has consequence on diabetes, heart problem and etc. With such lifestyle pattern, people health is being put at risk progressively. Although plenty of weight management tools are being offered, the rates of overweight and obesity still remains high. vTracker considers these mentioned issues with weight management tools by limiting the data entry process by giving the users the flexibility to choose the information from the drop-down list. This information is pre-maintained and stored in the database. Moreover, vTracker helps to consolidate the information of the workout with the dietary intake information using additional mobile auto workout tracker. A weekly report would be generated in order for the users to have the visibility of the dietary intakes and workout pattern. Other than that, weekly self-weigh in is also being included in the vTracker to let the users know whether they are on the right track or not. This service contributes to the success rate of the weight control [1]. Users can use vTracker online or through their mobile. With that, users can input the information at anywhere and anytime they wish. A reminder system is available to remind users about the data input.

In the following sections, UTAUT model is explained. Moreover, some of past studies with this model are stated. In section 3, a framework is proposed to predict the user's behavioural intention factors to use vTracker. Moreover, the meaning for the suggested framework's factors is defined in this section. In section 4, hypotheses are defined based on the suggested framework. In section 5, the research methodology is discussed. In section 6, the analyzed collected data from the questionnaires are presented. In section 7, a discussion regarding the result of the study is done. And finally, in last section, a conclusion is prepared based on the results from analyzing the hypotheses.

## 2. The unified theory of acceptance and use of technology model (utaut)

The Unified Theory of Acceptance and Use of Technology (UTAUT) model is modified from TAM. This model [2] is proposed by Viswanath Venkatesh et. all. on year 2003. The UTAUT model intends to interpret user intentions and usage behavior to use an information system. This model includes four variables. These variables are Performance Expectancy (PE), Effort Expectancy (EE) and Social Influence (SI). These variables are used to predict the user's behavioural intention to use a system. The fourth variable Facilitating Condition (FC) has direct influence towards user's behaviour. Moreover, four moderator variables are included in the model. These variables are gender, age, experience and voluntariness of use. They affect the strength of the relationship between dependent and independent variables in the model. The UTAUT model is shown in figure 1. In the following sections, a literature review about studies that consider UTAUT or extended UTAUT model as a research framework has been conducted.

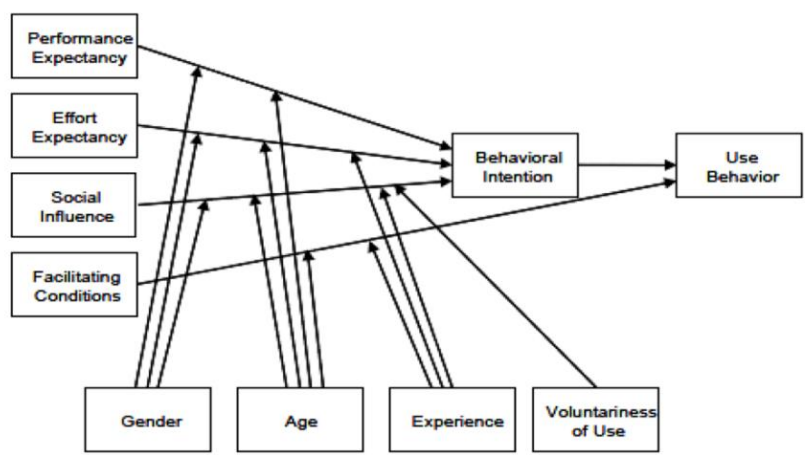

Figure 1: UTAUT Model [2]

## 3. The proposed research framework

The research framework of this study is adapted from UTAUT model [2] to measure the user's intention to adopt the vTracker. Due to time limitation, four moderator variables are not included in the study. Moreover, the Facilitating Condition (FC) factor is considered as a predictor for user's intention to adopt instead of user behavior. This is because vTracker is still at the conceptual stage where users can only be briefed on ideas of vTracker and Facilitating Condition (FC) cannot be tested on real products. The proposed framework is shown in Figure 2. Moreover, the definition of factors is shown in Table 1.

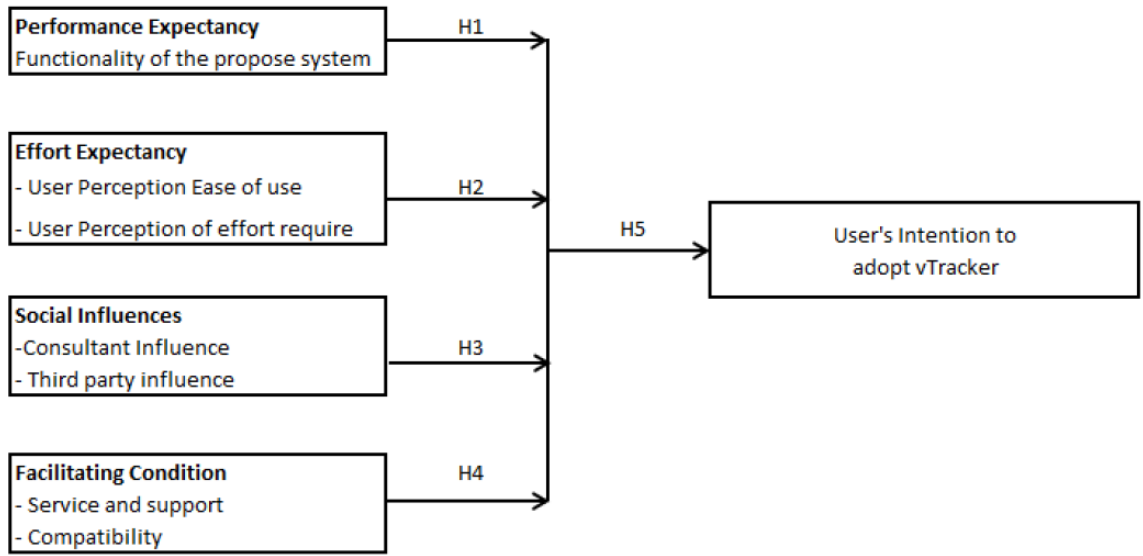

Figure 2: The proposed framework for vTracker

Table 1. The definition of factors in the proposed framework

| Variable | Definition |
|---|---|
| Performance Expectancy (PE) [2] | "The degree to which an individual believes that using the system will help him or her to attain gains in job performance." |
| Effort Expectancy (EE) [2] | "The degree of ease associated with the use of the system." |
| Social Influence (SI) [2] | "The degree to which an individual perceives that important others believe he or she should use the new system." |
| Facilitating Condition (FC) [2] | "The degree to which an individual believes that an organizational and technical infrastructure exists to support the use of the system". |
| User's intention [3] | "The degree to which a person has formulated conscious plans to perform or not perform some specified future behavior." |

## 4. Proposed hypothesis

In this study, specific hypotheses assumptions were proposed based on the previously UTAUT based models. The impact of these assumptions on user's intention to adopt VTracker among individual users in Malaysia was tested. In following, these assumptions are stated:

I. Performance expectancy of the system: A study conducted by by Foon, Yeoh Sok, and Benjamin Chan Yin Fah. shows that performance expectancy ($r = 0.51$, $p < 0.01$) are positively correlated with behavioral intention of the user [4]. Moreover, another study conducted by T. Escobar-Rodríguez and E. Carvajal-Trujillo shows that performance expectancy (B=0.099, $p<0.001$) positively impacts on behavioral intention of the user [5]. Therefore, if users believe that using vTracker could help them to have a reliable self-monitoring process, it would help them to achieve their goals of weight loss. Therefore, users would be more likely to adopt the vTracker for their dietary intakes monitoring purpose. With that, it can be determined that the higher the Performance Expectancy (PE) of user towards vTracker, the more likely vTracker would be adopted by the users. Research hypothesis (H1) is proposed as below:

H1: Performance Expectancy (PE) has a positive impact on the user's intention to adopt the vTracker.

II. Effort Expectancy of the system: vTracker concept is proposed to ease the process of data entry by providing a simple and user friendly data selection drop down list where user can easily select the information which they need rather than go through the hassle to figure out the amount of calories was consumed and burned throughout the workout activities. A study conducted by by Foon, Yeoh Sok, and Benjamin Chan Yin Fah. demonstrates that effort expectancy ($r = 0.55$, $p < 0.01$) is positively related with behavioral intention of the user [4]. Moreover, another study by T. Escobar-Rodríguez and E. Carvajal-Trujillo shows that effort expectancy (B=0.085, $p<0.01$) positively impacts on behavioral intention of the user [5]. Therefore, it is assumed that the higher the Effort Expectancy (EE) which user perceived the more likely they would adopt the vTracker. Therefore, research hypothesis (H2) is proposed as below:

H2: Effort Expectancy (EE) has a positive impact on the user's intention to adopt the vTracker.

III. Social Influence: A study conducted by by Foon, Yeoh Sok, and Benjamin Chan Yin Fah. Shows that social influence ($r = 0.64$, $p < 0.01$) is positively correlated with behavioral intention of the user [4]. Moreover, another study by T. Escobar-Rodríguez and E. Carvajal-Trujillo shows that social influence (B=0.043, $P<0.05$) positively influences on behavioral intention of the user [5]. Therefore, it is important for vTracker to have a good record towards user satisfaction so that the users recommend it to their relatives, friends or important person to them. Research hypothesis (H3) is developed as below:

H3: Social Influence (SI) has a positive impact on the user's intention to adopt the vTracker.

IV. Facilitating Condition: A study conducted by by Foon, Yeoh Sok, and Benjamin Chan Yin Fah. shows that facilitating condition (r = 0.63, p < 0.01) is positively correlated with behavioral intention of the user [4]. Moreover, another study by T. Escobar-Rodríguez and E. Carvajal-Trujillo shows that Facilitating Condition (B= 0.146, p<0.001) positively impacts on behavioral intention of the user [5]. Research hypothesis (H4) is developed as below:

H4: Facilitating Condition (FC) has a positive impact on the user's intention to adopt the vTracker.

## 5. Research methodology

The questionnaire's items were adopted from literatures which use UTAUT model to determine the user's intention to adopt online system or mobile applications. Moreover, the questionnaire's items were modified to suit vTracker context. The questionnaire was developed and distributed in hard and soft copy. Convenience sampling was used to reach through the respondents. The soft copy uses English language only but during the hard copy distribution, some of the respondents stated that they could not understand English well. Then the Malay version of the questionnaire was developed in order to cater to the respondents' needs since the target market is in Malaysia. For hard copy, a total of eighty questionnaires were distributed to friends, colleagues, relatives and their family members and fifty-two copies were returned. For soft copy, the questions for the questionnaire was developed and posted online at www.kwiksurvey.com. The respondents were notified about the online survey through email and social network websites like Facebook and Twitter. There were one hundred and seventy-six respondents participated in the survey but a total of twelve filled questionnaires were incomplete and being considered as void. This makes the total completed survey respondents to 216 people (164 softcopies and 52 hard copies) out of 350 participants.

## 6. Analysis and results

Reliability test is conducted on each items of the variables in the survey in order to study the consistency of the items and to find out the inter-relations among the survey items as a group. The result is shown as the Cronbach's Alpha value. In order for the survey items to be accepted as a group, the Cronbach's Alpha value needs to be 0.70 or higher. Once the Cronbach's Alpha value exceeds 0.7, it shows the higher the consistency of the collected data. The reliability test is very essential and needs to be completed before the other statistical tests as there are possibilities happen where some of the items need to be dropped in order to increase the consistency and reliability of the data. Table 2 shows the Cronbach's Alpha value for the three user's factor.

Table 2: Summary of Reliability Test

| Variables | Number Of items | Items Dropped | Cronbach's Alpha, α |
|---|---|---|---|
| Performance Effort Expectancy (PEE) | 6 | - | 0.841 |
| Social Influence (SI) | 4 | - | 0.909 |
| Facilitating Condition (FC) | 3 | - | 0.868 |
| Behavioural Intention to adopt (BI) | 3 | - | 0.893 |

From the results of reliability test, it can be concluded that no survey items are required to be dropped for the entire 3 user's factors as the Conbach's Alpha value are above 0.70. SI has the highest Conbach's Alpha value compare to the others factors with α = 0.909, followed by BI with α = 0.893, FC with α = 0.868 and finally PEE with α = 0.841.

Descriptive analysis is carried out to determine the level of user's intention to adopt vTracker. The data collections are based on 7-Likerts scale where 7- Strongly agree, 6- Agree, 5- Slight Agree, 4- Neutral, 3- Slight Disagree, 2- Disagree and 1- Strongly Disagree. From the results below, most of the factor's mean are at the range of 5 and it can be concluded that the respondents of the studies have the intention to adopt the vTracker as their self-monitored weight management tools. Table 3 shows the summary of descriptive analysis

Table 3: Summary of descriptive analysis

|  | Mean_PEE | Mean_SI | Mean_FC | Mean_BI |
|---|---|---|---|---|
| Mean | 5.449 | 4.8034 | 5.1246 | 5.1343 |
| Std. Deviation | 0.70555 | 1.03926 | 0.93826 | 1.01307 |
| Range | 3.67 | 5.25 | 5 | 5 |
| Minimum | 3.33 | 1.75 | 2 | 2 |
| Maximum | 7 | 7 | 7 | 7 |

Pearson correlation is used to test the relationship among each independent variable against dependent variable. Correlation in this context is defined as the degree of the variables related to another variable. The variables have positive relationship if Pearson correlation (r) is positive. Meanwhile, the variables are considered to have negative relationship, if the pearson correlation (r) is negative. From Table 4, the r value for PEE (r = 0.628), SI (r = 0.548) and FC (r=0.488) against BI is positive. Therefore, this result is in line with initial hypothesis construction where all the independent variables have a positive relationship with the dependent variable.

Table 4: Pearson correlation analysis results

|  |  | Mean_PEE | Mean_SI | Mean_FC | Mean_BI |
|---|---|---|---|---|---|
| Mean_PEE | Pearson Correlation (r) | 1 | .493** | .587** | .628** |
|  | Sig. (2-tailed) |  | 0 | 0 | 0 |
| Mean_SI | Pearson Correlation (r) | .493** | 1 | .437** | .548** |
|  | Sig. (2-tailed) | 0 |  | 0 | 0 |
| Mean_FC | Pearson Correlation (r) | .587** | .437** | 1 | .488** |
|  | Sig. (2-tailed) | 0 | 0 |  | 0 |
| Mean_BI | Pearson Correlation (r) | .628** | .548** | .488** | 1 |
|  | Sig. (2-tailed) | 0 | 0 | 0 |  |

**. Correlation is significant at the 0.01 level (2-tailed).

i. Regression Analysis

The simple regression analysis in this study is to find the relationship between the independent variables PEE, SI and FC against the dependent variable user's behavioural intention to adopt. From the results of the data analysis, it can be concluded that the PEE has the highest percentage towards the user's intention to adopt vTracker with the $r^2 = 0.394$ follow by SI with the $r^2 = 0.301$ and finally with FC with the $r^2 = 0.238$. All the 3 factors are also proven to be significant. The $r^2$ is a measure used to estimate the regression line against the real data point's line. The nearer the $r^2$ value towards 1, the better the model fits the real data which was collected. In other words, PEE is able to predict 39.4 percent of user's intention to adopt vTracker where SI and FC respectively can predict 30.1 percent and 23.8 percent on the user's intention to adopt vTracker. All the Beta value (β) are positive which means that PEE, SI and FC are directly proportional towards BI. For example, the higher the Beta value (β) for PEE factors in vTracker, the higher the user's intention towards adopting vTracker. Similar explanation implies towards the Beta value (β) of SI and FC where the higher the Beta value (β), the higher the user's intention towards adopting vTracker. The result of simple regression analysis is shown in Table 5.

Table 5: The result of simple regression analysis

| Independent Variable | Standardized Coefficient, Beta (β) | R value ( r ) | R Square (r2) |
|---|---|---|---|
| PEE | 0.628 | 0.628 | 0.394 |
| SI | 0.548 | 0.548 | 0.301 |
| FC | 0.488 | 0.488 | 0.238 |
| *p < 0.01 | | | |

From the Table 6, $r^2 = 0.478$ which means 47.8 percent of the data points fits our original proposed model. It can be concluded that PEE and SI is the most influential factors towards user's intention towards adopting vTracker with the p<0.01. Even though the p value for FC is more than 0.01, it is still less than 0.1. In statistical studies, the maximum p value is 0.1 which explains that confidence intervals which the data being accepted in a normal distribution. This implies that FC has influence on the user's intention to adopt vTracker but it is not as influential compare to PEE and SI. All the Beta values (β) are shown to be positive as well. The result of multiple regression analysis is shown in Table 6.

Table 6: The result of multiple regression analysis

| Independent Variable | Standardized Coefficient, Beta (β) | R value ( r ) | R Square (r2) | p value |
|---|---|---|---|---|
| PEE | 0.415 | 0.691 | 0.478 | p < 0.01*** |
| SI | 0.293 | | | p < 0.01*** |
| FC | 0.116 | | | p < 0.1* |

## 7. Discussion

From the descriptive analysis, the mean of all factors (PEE, SI and FC) is approximate to 5 and according to the 7- Likert scale, respondents of the research have agreed to adopt vTracker. From the multiple regression analysis results, the three factors have a direct positive relationship with the user's behavioural intention to adopt the usage vTracker. It is proven that all the three factors have positive impact on the user's intention to adopt vTracker while FC was being examined to have the least influence on affecting the user's intention to adopt vTracker compared to PEE and SI. A total of 47.8 percent of the data points fits our proposed model. This shows that the UTAUT model which was used to determine the user's intention to adopt vTracker is valid through the data collected.
All the three factors can be applied to the design of vTracker but the Performance Effort Expectancy (PEE) and Social Influence (SI) need to be emphasized compare to the Facilitaing Condition (FC).

## 8. Conclusion

The obesity and overweight issues are still on the rise in Malaysia even though there are lots of awareness campaigns to highlight this issue. These problems happen because Malaysians do not have the habit of self-monitor their dietary intakes and they also do not find overweight and obesity could bring them any harm. Moreover, the existing self-monitoring tools which are available in the market are not user friendly and cause a lot of inconveniences towards users in order to figure out the correct amount of calories intake and burned. This obstacle causes a major turn off for the users to use the existing self-monitoring tools. Therefore, vTracker is introduced with better features which would help to bring down the high dropout issues and maintain the consistency among the users. However, in order to make sure that this proposed solution work as planned, various models have been studied to understand the impact of user factors in influencing the user's intention to adopt vTracker. The Unified Theory of Acceptance and Use of Technology (UTAUT) are selected in order to predict the user factors which can influence users' intention to adopt vTracker. The UTAUT model consists of 4 dependent variables with 4 moderators. However, the four moderators are omitted due to time limitation. The four independent variables are Performance Expectancy (PE), Effort Expectancy (EE), Social Influence (SI) and Facilitating Condition (FC).
A survey questionnaire was developed and distributed among 206 respondents in Malaysia. Factor analysis results show that the initial model can be further improved by combing two factors PE and EE as one factor. The other two factors remain unchanged. Results show that respondents agree to use the vTracker. Moreover, all the three factors have positive relationship with user's intention behavioural to adopt vTracker. The user's factor which was determined would then be used to develop the prototype of vTracker because according to the basic concept of user acceptance model, the intentions to use the information technology would be a strong precursor to predict the actual usage of information technology.

**Acknowledgement**
The authors would like to thank Universiti Sains Malaysia (USM) as this research has been supported from the Research University Grant (RUI) [Account Number: 1001/PKOMP/811251] and from the Short Term Research Grant [Account Number: 304/PKOMP/6312103] from the Universiti Sains Malaysia.